\documentstyle[pra,aps,epsfig]{revtex}
\begin{document}
\draft
\title{The hyperfine transition in light muonic atoms of odd Z}
\author{T.J. Stocki,\thanks{email: trevor.stocki@crc.ca}
\thanks{Present address:
Communications Research Centre, 3701 Carling Avenue, Box 11490, Station H,
Ottawa, Ontario, Canada, K2H 8S2.}
D.F. Measday,
E. Gete,\thanks{Present address: 
London Regional Cancer Centre, 790 Commissioner Road East, London, Ontario,
Canada, N6A 4L6.}
M.A. Saliba,\thanks{Present address:
Faculty of Engineering, University of Malta, Msida  MSD 06, Malta.}
J. Lange,\thanks{Present address:
Defence Research Establishment Ottawa, 3701 Carling Avenue,  Ottawa, Ontario,
Canada, K1A 0Z4.}
}

\address{Department of Physics and Astronomy \\
         University of British Columbia \\
         Vancouver, B.C., Canada V6T 1Z1}

\author{T.P. Gorringe}
\address{Department of Physics and Astronomy \\
         University of Kentucky \\
         Lexington, KY, U.S.A. 40506-0055}

\date{\today}
\maketitle
\begin{abstract}
   The hyperfine (hf) transition rates for muonic atoms have been re-measured
for select light
nuclei, using neutron detectors to evaluate the time dependence of muon 
capture.  For $^{19}$F $\Lambda$$_{h}$ = 5.6 (2) $\mu$s$^{-1}$ for the hf
transition rate, a value which is considerably more accurate than previous
measurements.  Results are also reported for Na, Al, P, Cl, and K; that result
for P is the first positive identification.
\end{abstract}

\section{Introduction}
\label{}
There is still much uncertainty about the hyperfine transition rate for muonic
atoms in the 1s ground state.  Different techniques often give different rates,
and there are many inconsistencies.  This topic was boosted into prominence 30
years ago by the Chicago group of Winston, Telegdi and co-workers \cite{r1,r2}
who studied $^{19}$F and established  the major properties of this effect.
Even today, $^{19}$F remains the best understood example, because of the
convenient time constant of about 180 ns. \\

When a $\mu$$^{-}$ stops in a target, it forms a muonic atom, and quickly 
cascades down to the 1s level on a time scale of ~10$^{-12}$ s which is 
effectively instantaneous for any detector.  If the nucleus has no spin,
there is a single ground state and the muon awaits its fate via the weak
interaction decay
($\mu$$^{-}$ $\rightarrow$  e$^{-}$ $\overline{\nu}_{e}$  $\nu$$_{\mu}$ )
or via nuclear
capture.  For the light elements this occurs within a few $\mu$s and can
be studied with a variety of detectors.  If the nucleus has a non-zero spin 
however, the situation
is more complicated because there are two hyperfine levels of the 1s state,
separated by an energy varying between a few eV and a keV or so.  If the
transition between these levels occurred via an M1 photon, the rate is
too slow to be observed. However, the Chicago group showed that Auger emission
can
speed up the transition rate a thousand fold and bring it to the time scale of
nanoseconds, which is within the range of standard detectors.  The hyperfine
transition
can be detected in any nucleus by observing the depolarization of the
$\mu$$^{-}$ via the detection of its decay electron.  However, the $\mu$$^{-}$
has a very small residual polarization in the atomic state ($<$ 10\% for nuclei
with spin), and in addition the $\mu$$^{-}$  can be depolarized by magnetic
interactions or by nearby radicals.  Nevertheless, several experiments have
been carried out successfully.  \\

If the nucleus has an odd Z, the $\mu$$^{-}$ is magnetically coupled to the 
odd proton.  Now the capture probability in the $\mu$$^{-}$p system, is 660
s$^{-1}$ for the singlet state, but only ~12 s$^{-1}$ for the triplet state.
Thus in one hyperfine state the total muon capture probability is
approximately proportional to (1 + Z/2)
and in the other proportional to (Z/2).  Thus nuclear capture is highly 
sensitive to the hyperfine state, and the hyperfine transition can be followed
by detecting neutrons from the capture events.  The effect can also be observed
via the time-dependence of the decay electron, but in light nuclei the normal
decay rate dominates and thus dilutes the signal.  Winston \cite{r2} as well as
Suzuki et al. \cite{r3} were able to observe this effect in $^{19}$F, but for
other nuclei the signal is too small. \\

The hyperfine transition can also be observed by detecting specific
$\gamma$-rays resulting from nuclear capture.  Some transitions are highly
sensitive to the initial hyperfine state and, for certain spin combinations,
this is also true for even-Z nuclei such as $^{13}$C.  Several nuclei have been
studied by Gorringe and coworkers
\cite{r4,r5,r6,r7}.  In addition Wiaux \cite{r8} has found a very large 
sensitivity for the 320 keV $\gamma$-ray from muon capture in $^{11}$B.  An
unexpected observation by Gorringe et al., was that the hyperfine rate in
metallic sodium was $\Lambda$$_{h}$ = 15.5 $\pm$ 1.1 $\mu$s$^{-1}$ \cite{r5},
but when in NaF, the sodium hyperfine rate was only 
$\Lambda$$_{h}$ = 8.4 $\pm$ 1.9 $\mu$s$^{-1}$ \cite{r4}; the fluorine
hyperfine rate was $\Lambda$$_{h}$ = 4.9 $\pm$ 1.2 $\mu$s$^{-1}$, consistent
with Winston's measurement of $\Lambda$$_{h}$ = 6.1 $\pm$ 0.7 $\mu$s$^{-1}$
\cite{r2}, and our own measurement of
$\Lambda$$_{h}$ = 5.6 $\pm$ 0.2 $\mu$s$^{-1}$ in
LiF.  This raised the possibility that there might be a
difference between metals and insulators, or between different molecular
species.  Such a difference could be due to different recombination times, or
an unusual
sensitivity to the binding energy of the valence electrons.  However both 
possibilities seem unlikely. \\

The most puzzling result to date has been the observation of a hyperfine 
transition in $^{14}$N by Ishida et al. \cite{r9}, using the $\mu^{-}$SR
technique
to observe the muon depolarization. They detected a relaxation rate of
0.092(33) $\mu$s$^{-1}$ of which about 0.016 $\mu$s$^{-1}$ is due
to the different loss rates from the hyperfine states.  This leaves 0.076(33) 
$\mu$s$^{-1}$ 
which may be from a hyperfine transition.  The problem is that the energy
difference between the hyperfine states is only 7.4 eV, whereas, in the carbon
atom, the least bound electron is bound by 11.3 eV, i.e. this energy is the
ionization potential (the $\mu$$^{-}$ is close to the nucleus, so the $^{14}$N
muonic system appears to the electrons to be more like a $^{14}$C nucleus).  If
the pseudo $^{14}$C atom has formed a ``C''N bond, the ionization potential
is 14.3 eV, which makes the matter worse.  We note that Wiaux found a slower
hyperfine rate in $^{11}$B in comparision to a $\mu$SR experiment, so we
suggest that the observation in $^{14}$N was caused by an additional
depolarization mechanism, not the hyperfine transition. \\

Because of the various inconsistencies in this field, it was decided to
re-investigate several light nuclei, using the more reliable technique of 
neutron detection.  A comparison was also desirable between Na and NaH,
Al and LiAlH$_{4}$, and K and KH, to see if any difference could be detected.  
Unfortunately, no satisfactory result was obtained for KH.
It should be noted that any muon captured in hydrogen is immediately
transferred to another atom
because the $\mu$p system is neutral.  Thus NaH, for example, is effecively
pure sodium in an insulating environment. \\

\section{Experimental Technique}

The experiment was carried out in the M9B channel at TRIUMF, which includes
a superconducting solenoid.  For muons of 60 MeV/c the stop rate was $\approx$ 
10$^{4}$ s$^{-1}$ with an electron contamination of 2 \% and a pion
 contamination of 
$\leq$ 0.2 \%.  The muons passed through 2 plastic scintillators, and stopped
in various targets; a third large plastic scintillator was used as a veto to
define a muon stopped in the target.  A mu-metal shield around the targets
reduced the 
ambient magnetic field from 1.5 to 0.1 gauss. Table~\ref{tab1} lists the
properties of the targets.\\

The neutron detectors were four cylindrical liquid scintillators, 2 of NE213,
1 of NE224 and 1 of BC501A (equivalent to NE213).  They were arranged in a 
symmetrical array at 45$^{\circ}$, 135$^{\circ}$, 225$^{\circ}$, and
315$^{\circ}$ to the beam, in order to minimize muon rotation effects. Plastic
scintillators were placed in front of each detector to veto charged particles
(such as decay electrons).  The counters had a timing resolution of better
than 5 ns full width half maximum (FWHM) for $\gamma$-rays.  The counters
were about 30 cm from the target so a $\gamma$-ray time of flight is 1 ns. \\

Pulse shape discrimination was used to distinguish neutrons from
$\gamma$-rays.  Two different modules were used at different times, but the
discrimination was always set conservatively \cite{r10}.  Sources of $^{60}$Co
and AmBe
were used to set up the system.  If $\gamma$-rays are detected during the
experiment, a prompt
peak is clearly distinguishable, so it is straightforward to monitor the 
electronics on-line.  \\

The time of arrival of an event with respect to the stopping muon was measured
by both a 5 $\mu$s and a 10 $\mu$s
full scale time to digital converter (TDC).  These TDCs have been tested and
do not contribute significantly to the errors in the experiment. \\

Data acquisition was carried out by a VAX station 3200 and a PDP-11
front\-end
processor (starburst).  Data collected from the CAMAC modules were written
to 8 mm tapes for later analysis.  Further technical details of the equipment
are available in the thesis by Stocki \cite{r10}.  \\

To test the overall system several targets were chosen for which there is no 
hyperfine transition, or for which the transition is too fast to be observed.  
The time of arrival of the neutrons could then be fitted by the formula:

\begin{equation}
N(t) =  \left( \frac{1}{2} erf \frac{t-t_0}{\sqrt{2} S} + \frac{1}{2} 
\right) C e^{-Dt} + B \label{e1}
\end{equation}

where D is the total disappearence rate for the element under study, B is a
flat background,
t$_{0}$ is the mid-point of the rise-time curve, and S is a folding of the time
 of flight spread and instrumental time resolution. \\

\begin{figure}
\begin{centering}
\epsfig{figure=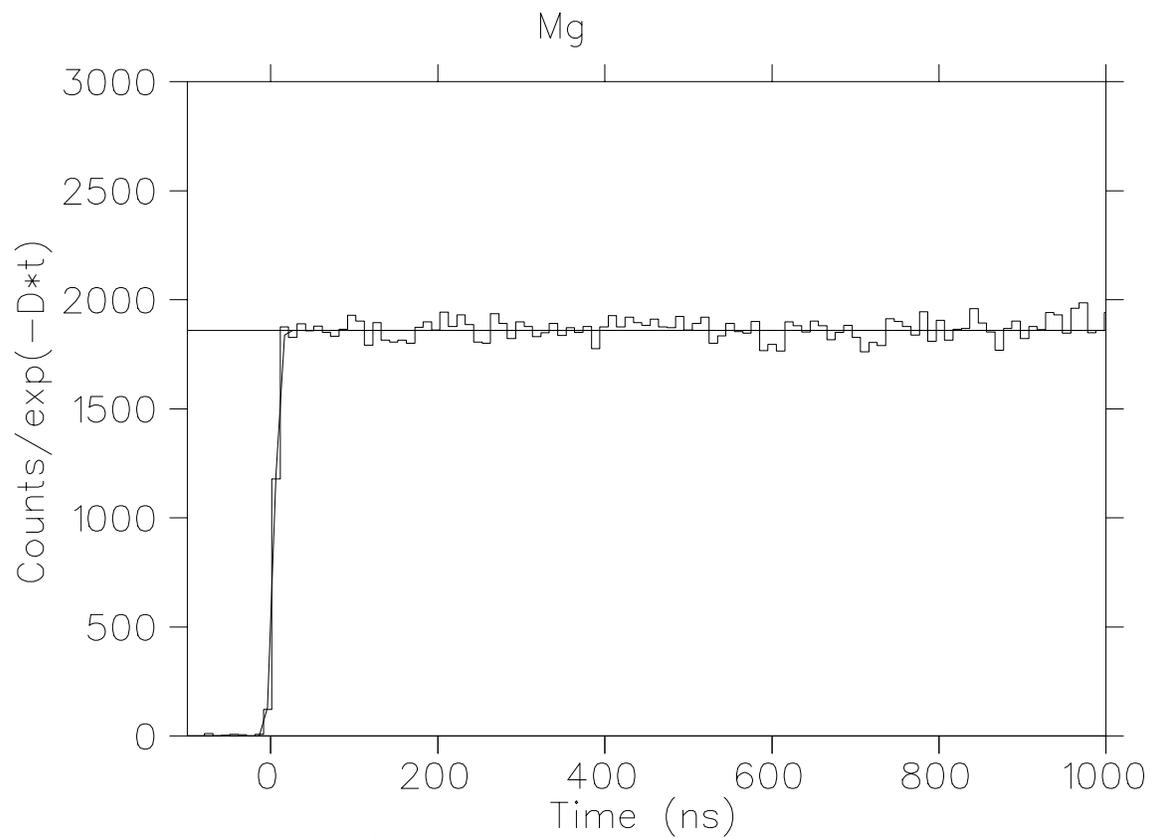,height=15cm,angle=90}
 \caption{The time spectrum of neutrons from muon capture in Mg, fitted to
 equation 1.  The data are presented with the muon disappearance rate divided
out.}
\label{fig1}
\end{centering}
\end{figure}

A typical fit for Mg is illustrated in Figure~\ref{fig1}. The values for S
 and t$_{0}$ are presented in Table~\ref{tab2} 
and show the
tendency for heavier elements to have a slower time of flight for the neutrons.
  As t$_{0}$ refers to the time of arrival of the $\gamma$-rays, one can
add the $\gamma$-ray time of flight (0.97 ns) to obtain the average neutron
time of flight ($\Delta$t) and average neutron energy ($\overline{E}$).  From
S one can remove the counter
timing resolution to obtain the spread in energy of the neutrons ($\Delta$E).  
These values were used as guidance in analyzing the elements with a hyperfine
effect. Note that these values for $\overline{E}$ and $\Delta$E are for the
neutrons detected in this experiment, and do not constitute measurements of
these parameters for muon capture in general.\\

\begin{figure}
\begin{centering}
\epsfig{figure=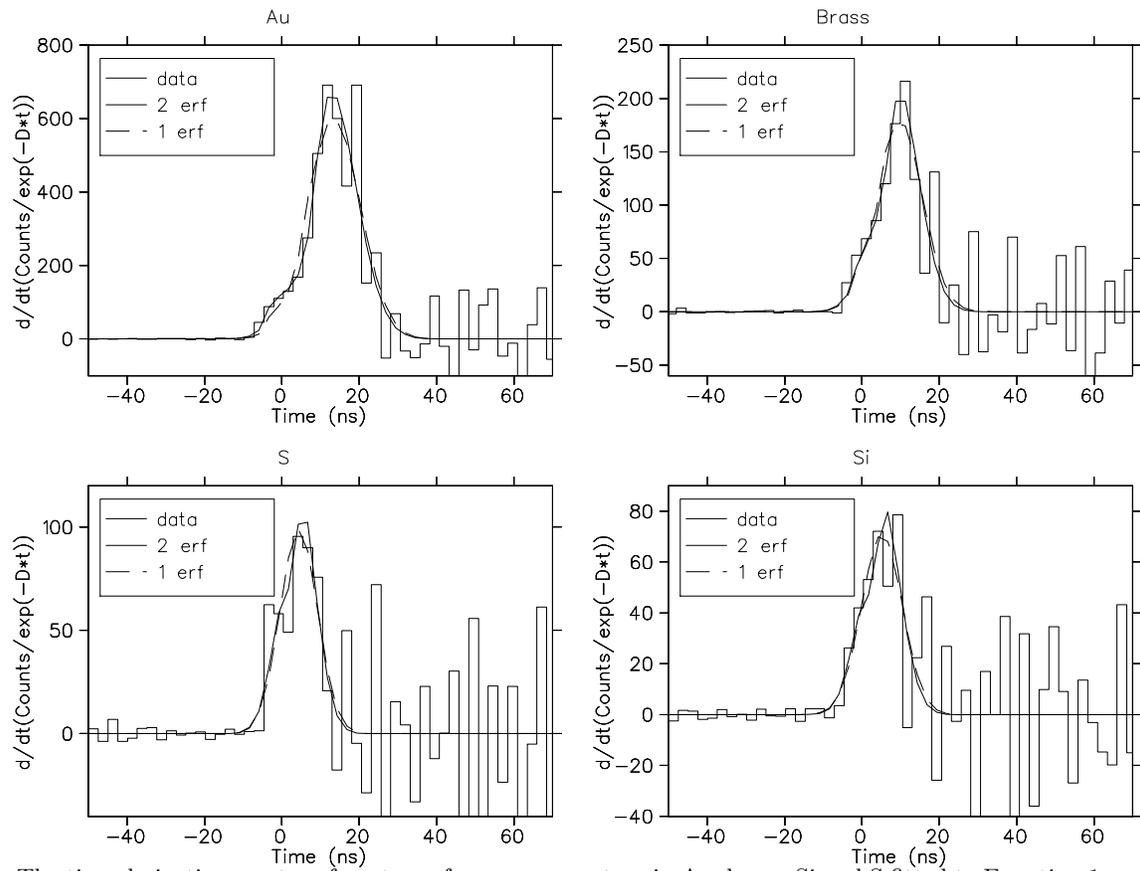,height=15cm,angle=90}
 \caption{The time derivative spectra of neutrons from muon capture in Au, 
brass, Si and S fitted to Equation 1 and Equation 2.}
\label{fig0}
\end{centering}
\end{figure}

By taking the time derivative of the data such as that shown in 
Figure~\ref{fig1}, one can get an idea of the neutron
time of flight and detector resolution effects.  Figure~\ref{fig0} shows this
time derivative spectra for Au, brass, Si, and S.  One can clearly see in
brass and Au that a single error function (or gaussian) is not enough.  We
found that the function:

\begin{equation}
N(t) =  \left[\frac{1}{2} erf \frac{t-t_0}{\sqrt{2} S} + \frac{1}{2} + R\left(
\frac{1}{2} erf \frac{t-t_1}{\sqrt{2} S_1} + \frac{1}{2} \right) \right]C
e^{-Dt} +  B \label{e2}
\end{equation}  

seems to give a better fit for those elements.  The effect is not seen in Si
and S, due to limited statistics. \\

For most elements there was a small correction (a few percent) for background
from muons stopping in carbon in the scintillators.  In addition, a few targets
contained two elements viz LiF, (CF$_{2}$)$_{n}$, LiCl, and CCl$_{4}$. 
We used teflon, i.e.\,(CF$_{2}$)$_{n}$, and CCl$_{4}$ as test cases to ensure
that the
corrections could be adequately applied.  For such molecular targets one needs
to know the atomic capture ratio.  The Fermi-Teller Z law is far from adequate
and a Z$^{\frac{1}{3}}$ variation gives a better approximation \cite{r11}. 
However, far more
satisfactory is to use an actual measurement as the empirical variation is 
significant. \\

For LiF the atomic capture ratio (Li/F) has been measured to be 0.28(3) by
Zinov et al. \cite{r12} and 0.10(8) by Wilhelm et al. \cite{r13} for muons,
and 0.10(1) \cite{r14} and 0.22(2) \cite{r15} for pions, which should be the
same.  As Zinov et al. disagree with later measurements for several other
molecules, we take the pion values and use 0.16(12) with a liberal error.  This
value agrees with the empirical capture probability for muons determined by
von Egidy and Hartmann which is 0.18(5) for LiF \cite{r16}.  For
(CF$_{2}$)$_{n}$, Martoff et al. \cite{r15} found that (89.4 $\pm$ 0.7) \%
of pions capture on fluorine (i.e. an atomic capture ratio C/F of 0.24(2) in
the normal definition, which takes into account the relative number of atoms.
(The Z$^{\frac{1}{3}}$ law gives 0.87; von Egidy and Hartmann \cite{r16} do
not give a value for carbon, but using the Martoff value, and a measurement
of CO$_{2}$ \cite{r17}, we estimate the C capture probability is 0.33(9) in 
the von Egidy-Hartmann scheme).  For CCl$_{4}$ there is no measurement but
using our value of the C capture probability we obtain C/Cl = 0.25(7) i.e.
(5.8 $\pm$ 1.6) \% of muons stop in the carbon of CCl$_{4}$.  The
Z$^{\frac{1}{3}}$ rule gives C/Cl = 0.71, i.e. 15 \% of muons stop in the
carbon
of CCl$_{4}$.  For LiCl the atomic capture ratio, as measured by Daniel et al.
\cite{r18} is 0.19(8). \\

Although these values are very uncertain, the situation is not as bad as
it might appear.  Since the lighter elements produce many fewer neutrons,
the corrections turn out to be fairly minor.  Hence, the number of stops in an
element for which capture occurs (most of which produce neutrons), are 0.53\%
for Li, 7.8\% for C, 33 \% for F and 75 \% for Cl \cite{r3}.  The effective
background, therefore, from lithium in LiF is 0.25 \%, which is negligible
in comparison to other problems.  We assume the neutron multiplicity to be the
same for these elements.

\section{Experimental Results}
\begin{figure}
\begin{centering}
\epsfig{figure=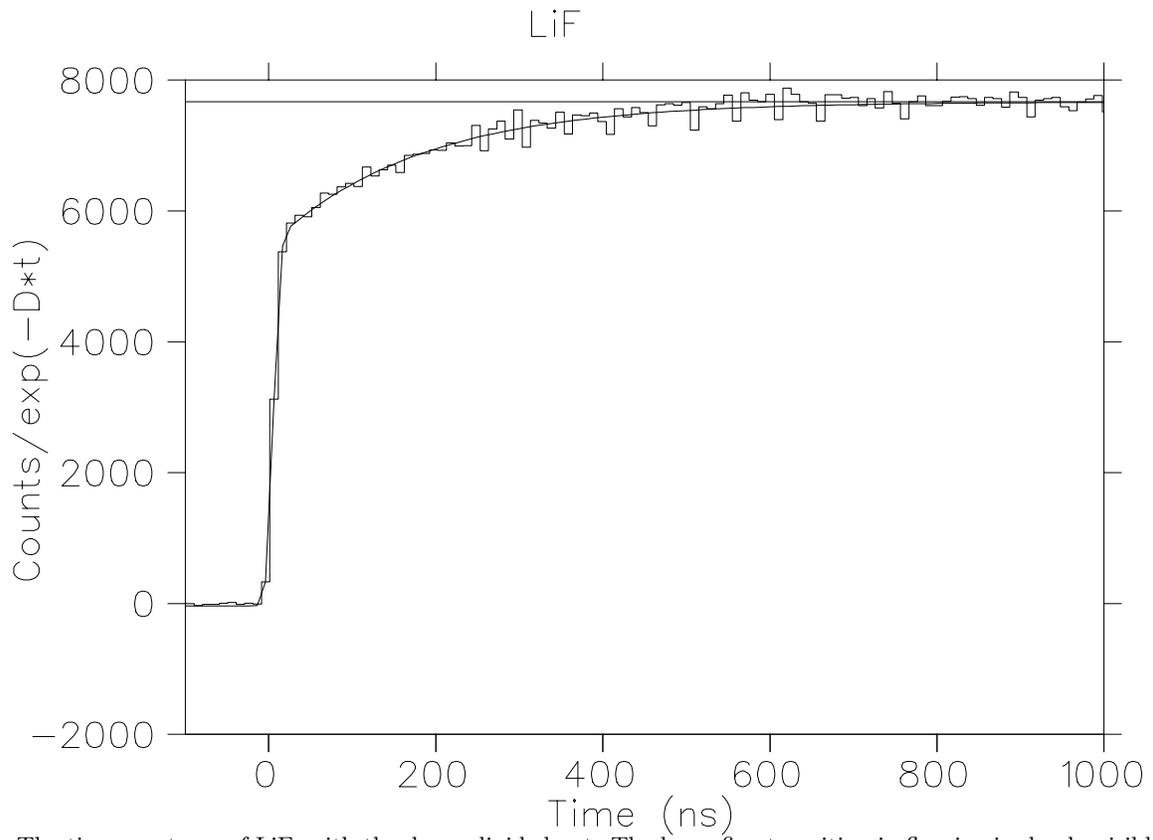,height=15cm,angle=90}
 \caption{The time spectrum of LiF, with the decay divided out. The hyperfine
transition in fluorine is clearly visible with its time constant of about 180
ns.}
\label{fig2}
\end{centering}
\end{figure}

The results for elements with a hyperfine effect are fitted to two different
functions; one with a single error function viz

\begin{equation}
N(t) =  \left[ erf \frac{t-t_0}{\sqrt{2} S} + 1 \right]\left[C e^{-Dt} \left(1
- A e^{-Ht} \right)  + F e^{-Gt} +  P e^{-Qt} \right] + B \label{e3}
\end{equation}

and another more complex function with two error functions:
\begin{eqnarray}
N(t) & = & \left[ \frac{1}{2} erf \frac{t-t_0}{\sqrt{2} S} + \frac{1}{2} +
 R\left(\frac{1}{2} erf \frac{t-t_1}{\sqrt{2} S_1}  + \frac{1}{2} \right)
 \right] [ C e^{-Dt} \left( 1 - A e^{-Ht} \right)  \nonumber \\
 & & \nonumber \\
 & & + F e^{-Gt} +  P e^{-Qt} ] + B \label{e4}
\end{eqnarray}

where H is the hyperfine rate, A is the hyperfine asymmetry, F and G are
contributions from carbon in the counters, P and Q are contributions from other
elements (if appropriate), and B is a flat background.  The parameters F, G, P,
and Q were calculated and fixed.  These components make no substantial 
contribution to the data analysis. By using these two equations we can get a
handle on the systematic errors.  In Equation 4 we tie the S$_1$ and t$_1$
values to the S and t$_0$ for two cases, using fits to the brass and the Au
data.  For LiF the data were of sufficient 
quality that all the other paramters (t$_{0}$, S, C, D, A, H) could be fitted
freely.  In this case the hyperfine effect is well separated from the rise-time
effect caused by the neutron time of flight effects.  The best fit is
illustrated in Figure~\ref{fig2} and the values of S and t$_{0}$ are given in
Table~\ref{tab3}.
They are consistent with the non-hyperfine elements but slightly different.\\

This is a salutary warning because we shall need to
use both Equations (3) and (4) to get a handle on the systematic errors due
to the time of flight and timing resolution effects.  Our recommended values
are the averages of these two methods.  One would think that one could get
the values of S and t$_{0}$ from the time spectra of the neighbouring 0$^{+}$
nuclei.  This turns out to be a bad assumption.  First let us understand the
physics behind this hypothesis.  Nuclei
without a hyperfine effect are 0$^{+}$ nuclei, often even-even nuclei such as
$^{16}$O, $^{24}$Mg, $^{28}$Si, and $^{40}$Ca.  These are tightly bound.
However odd Z nuclei like $^{19}$F, $^{23}$Na, $^{27}$Al, and $^{31}$P
tend to have N=Z+1 and so, transforming a proton to a neutron proceeds to a
nucleus even more neutron rich and more likely to fall apart.  The typical 
neutron energy spectrum after muon capture is composed of two components, an
evaporation spectrum peaked at about 1.5 MeV followed by a high energy tail
starting around 5 or 6 MeV \cite{r19}.  Setting a threshold at 10 MeV 
Kozlowski et al. \cite{r20} have shown that high energy neutrons constitute
the following fraction: 26(5)\% in $^{16}$O, 19(3)\% in $^{28}$Si,  11(2)\% in
Ca and 11(2)\% in Pb.  Thus for light elements the high energy component is
very important, and it is not surprising if it is sensitive to details of
the nuclear structure of the product nucleus.  The results of our experiment
confirm that there are significant variations from nucleus to nucleus. \\

\begin{figure}
\begin{centering}
\epsfig{figure=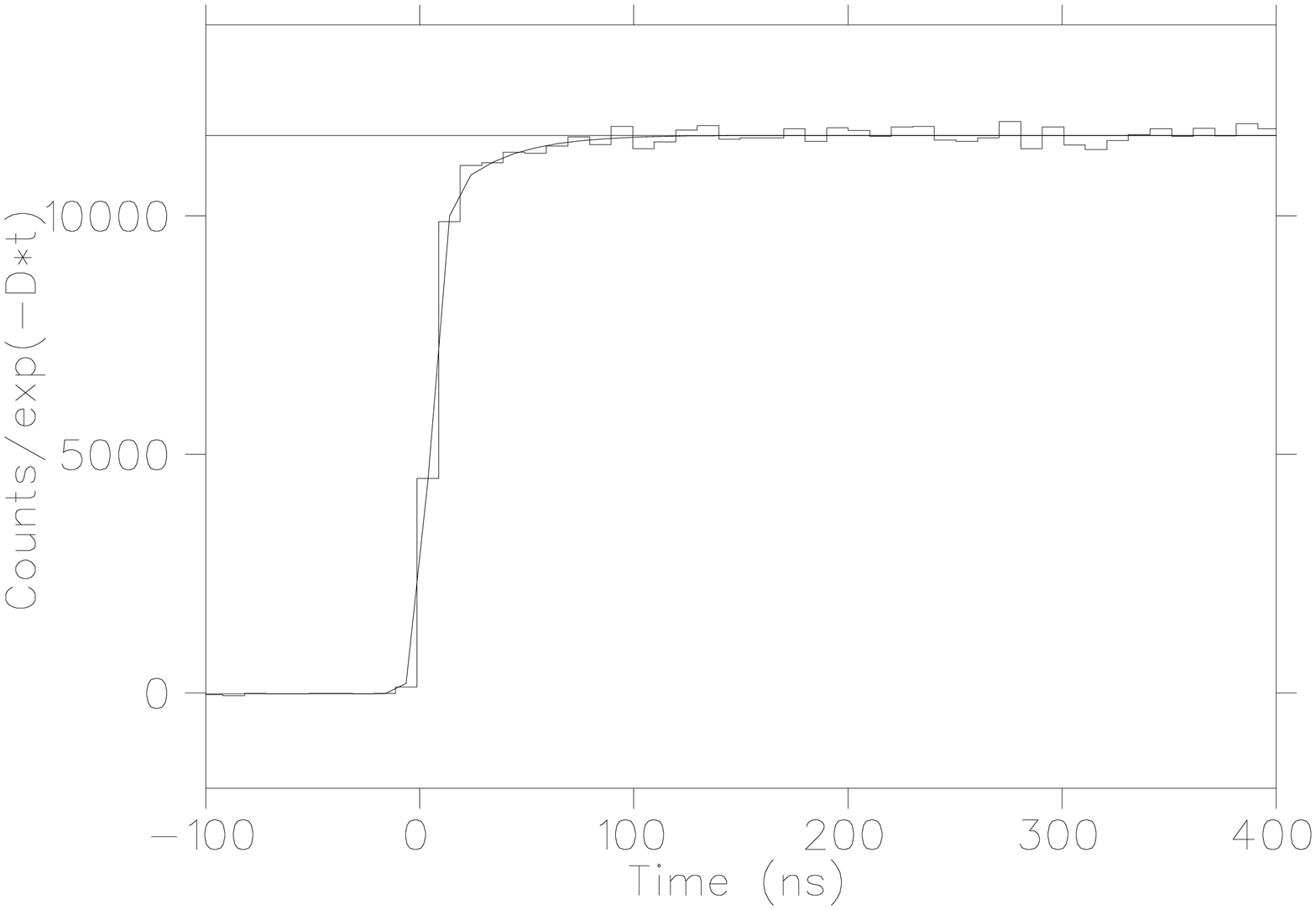,height=15cm,angle=90}
\epsfig{figure=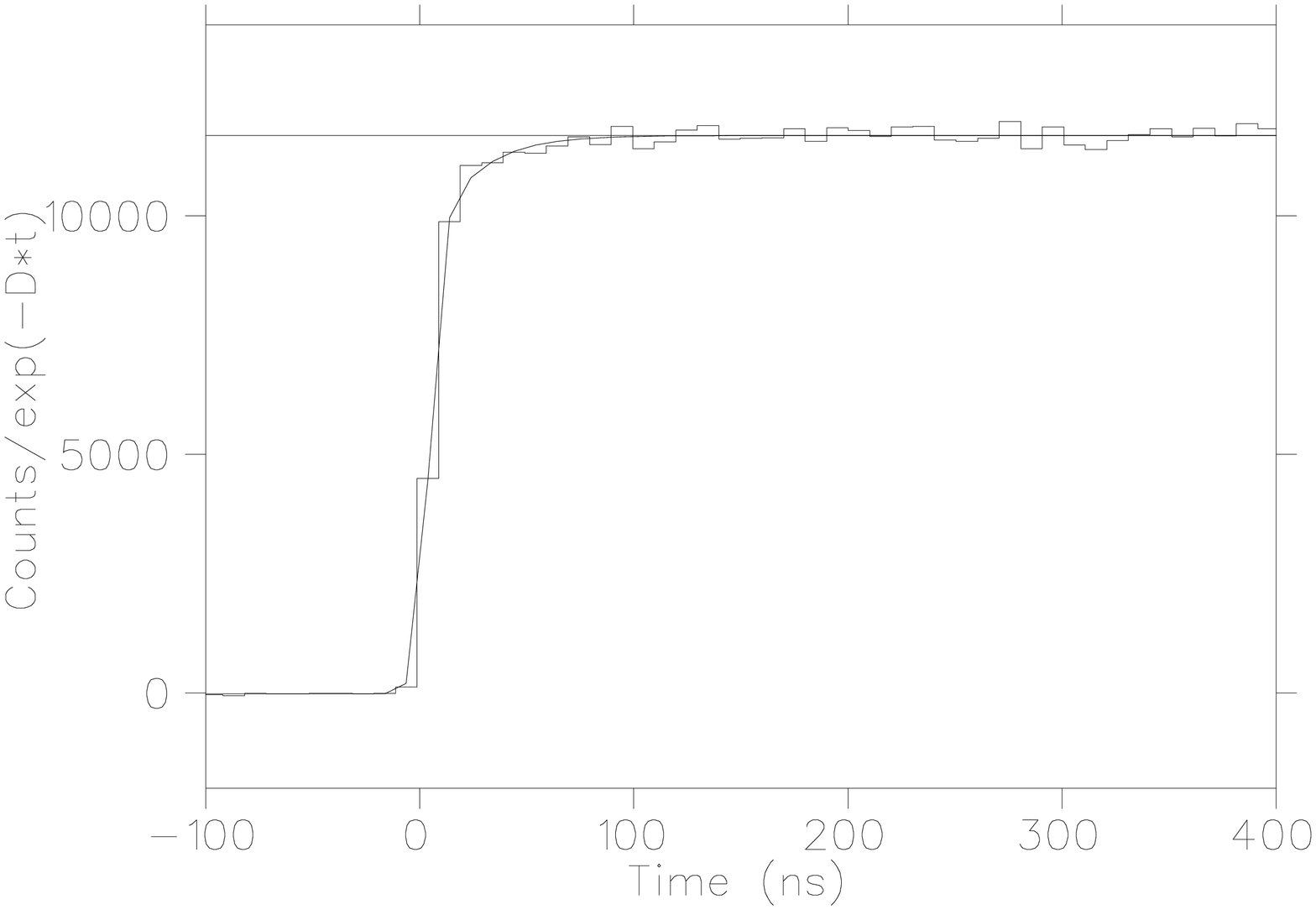,height=15cm,angle=90}
 \caption{The time spectrum of Na, with the decay divided out for the one
error function case (top) and the two error function case (bottom).}
\label{fig3}
\end{centering}
\end{figure}

For Na, Al, P, Cl, and K we do not have sufficient data to allow a free
parameter search, so we must average the single error function results with
the two error function results, otherwise we do not know whether the results
are from both time of flight and hyperfine effects.  Notice for LiF and
(CF$_{2}$)$_{n}$ this does not have to be done, because in F the capture rate
asymmetry is large enough and the hyperfine rate is slow enough that the
hyperfine rate can be distinguished from the time of flight effects.  One way
out of the dilemma for the other elements is to raise the energy 
threshold on the neutron detector.  However, the hyperfine asymmetries are
very small, and statistics becomes a problem.  The other way is to do a
full study of the neutron spectrum but this is a major project.  We have
therefore been forced to simply compare results from different fitting 
functions, using shape information from the spectra in Figure 2. 
The difference between the fits gives
us an estimate of the systematic error.  In Table~\ref{tab3} we present some
 of these
fits to illustrate the problem; Figure~\ref{fig3} shows two fits
 for Na which are both quite satisfactory.
In Table~\ref{tab4} our final recommended values are compared with previous
results and the 
hyperfine asymmetries are compared in Table~\ref{tab5} to various theories,
and previous
experiments.  Note that in Table~\ref{tab4} the Cl and K results did not
use these error functions; since their asymmetry values were $<$ 0, the
hyperfine effect could be easily distinguished from the neutron time of flight
effects.  The most important feature is that apart from $^{19}$F and to 
some extent $^{23}$Na, the hyperfine asymmetries are small, and that creates
part of the problem.  The $\gamma$-ray experiments observe large asymmetries
(and $\gamma$-rays all travel at the same speed).  Thus $\gamma$-ray
experiments
with lower statistics can effectively compete with neutron-detection
experiments.  A word of caution should be given concerning the comparison of
the asymmetry with previous calculations, which were made for the
total capture rates (including transitions to bound levels which give off no
neutrons, and to levels which give off 2 neutrons).  The experimental
asymmetries are for specific experimental conditions with a threshold on the
neutron energy and an energy dependent efficiency for the neutron detector.
Thus one would not expect an exact equivalence, but the comparison is 
interesting. \\ 

We see that, on the whole, the errors on the hyperfine rate are somewhat
large.  The two results which are most convincing are F and Na, but the latter
is not sufficient to resolve the disagreement between the earlier sodium metal,
and NaF results. \\

The phosphorus result is worth a short discussion.  An initial polarization 
measurement by Egorov et al. \cite{r23} claimed that an asymmetry had been
observed.  The measurement was from 0 to 2.4 $\mu$s implying no fast
depolarization.  Note also that the upper level is F=1 and the lower level
F=0, so that if a fast hyperfine transition does occur, there can be no
residual
polarization.  Later asymmetry measurements by Hutchinson et al. \cite{r24} 
and Babaev et al.\cite{r25} observed no asymmetry and set limits at 10\%
of the value observed by Egorov et al., thus indicating that the hyperfine
transition was fast, as suggested by Winston \cite{r2} who calculated 58
 $\mu$s$^{-1}$, i.e. $\tau$ = 17 ns.  This scenario was consistent with
the data of Gorringe et al. \cite{r5} using $\gamma$-ray detection, but no
positive observation was obtained.  Again, our own result has systematic
difficulties but is the first positive identification of the hyperfine
transition in phosphorus.  All the other hyperfine rates are consistent with
earlier results. \\

The comparision of Na and NaH in our data is interesting, but not of sufficient
quality to resolve the questions about the sodium hyperfine rate, similarly 
for Al and LiAlH$_{4}$, though the trend is for the hyperfine rates to be 
similar. \\

In conclusion, we have measured the hyperfine rate in P for the first time,
and obtained a more accurate value for F.  Even though neutron detection
appears to be the most sensitive technique for measuring the hyperfine
transition rate, the problems with the neutron spectrum and the spread of the
time of flight to the detectors have made it very difficult to observe fast
rates.  Further experiments should use a higher energy threshold (and take
considerably greater statistics).  An alternative technique would be to use a
$\gamma$-ray detector with a high resolution, but a faster time response than
a HPGe detector.

\acknowledgements
We wish to thank Glen Marshall and Elie Korkmaz for lending some liquid
scintillators and some pulse shape discriminator NIM modules to us.  We would
also like to thank B.A. Moftah and M.C. Fujiwara for their help and advice
with the experiment.  We thank Lars Holm for his help with the University of
Alberta pulse shape discrimination NIM module.  We wish to thank the Natural
Sciences and Engineering Research Council of Canada and similarly the National
Science Foundation in the U.S.A. for providing support and
equipment for this experiment; we also thank the National Research Council of
Canada and the staff at TRIUMF for providing the excellent muon beam facility.


%
%

\pagebreak
%
%

\begin{table}[t]
\begin{centering}
 \caption{Properties of the targets used. }
\begin{tabular}{ccdd}  
Z & Target & Thickness & Number of Events \\
 & & (g/cm$^{2}$) & in Time Spectrum \\ \hline 
3,9  & LiF & 2.70 & 1.2 $\times$ 10$^6$ \\
6,9  & (CF$_{2}$)$_n$ & 2.25 & 3.5 $\times$ 10$^5$ \\
11  & NaH & 1.14  & 1.5 $\times$ 10$^6$ \\
11  & Na & 1.65  & 1.5 $\times$ 10$^6$  \\
12  & Mg & 1.61 & 2.2 $\times$ 10$^5$  \\
13  & Al & 4.69  & 1.2 $\times$ 10$^6$  \\
13  & LiAlH$_{4}$ & 1.51  & 1.9 $\times$ 10$^5$  \\
14  & Si & 3.55  & 3.0 $\times$ 10$^5$  \\
15  & P & 2.79 & 2.8 $\times$ 10$^6$  \\
17  & LiCl & 2.60  & 7.6 $\times$ 10$^5$  \\
17  & CCl$_{4}$ & 6.18 & 4.9 $\times$ 10$^5$ \\
19  & K & 1.78  & 8.4 $\times$ 10$^5$  \\
29,30  & Brass (Cu+Zn) & 2.79 &  1.8 $\times$ 10$^5$  \\
\end{tabular}
\label{tab1}
\end{centering}
\end{table}

\begin{table}[t]
\begin{centering}
 \caption{Values of S and t$_{0}$ from non-hyperfine targets.  Also given
are the average energy $\overline{E}$, and the energy spread $\Delta$E of the
neutron events used for the timing measurements. }
\begin{tabular}{cc|r@{}l@{}r@{}lr@{}l@{}r@{}l|r@{}l@{}r@{}lr@{}l@{}r@{}l
r@{}l@{}r@{}l}
Z & Target & \multicolumn{4}{c}{S} & \multicolumn{4}{c|}{t$_{0}$} &
\multicolumn{4}{c}{$\Delta$t} &\multicolumn{4}{c}{$\overline{E}$ (MeV)} &
\multicolumn{4}{c}{$\Delta$ E} \\
 & & \multicolumn{4}{c}{(ns)}  & \multicolumn{4}{c|}{(ns)} &
 \multicolumn{4}{c}{(ns)} & \multicolumn{4}{c}{(MeV)} &
 \multicolumn{4}{c}{(MeV)}  \\ \hline 
12 & Mg  & 4&.63\,\, & $\pm$ 0&.24  & 4&.22\,\, &  $\pm$ 0&.23  & 5&.19\,\, &
 $\pm$ 0&.23 & 16&.5\,\, & $\pm$ 1&.6 & 27&\,\, & $\pm$ 4&  \\ 
14 & Si  & 5&.26\,\, & $\pm$ 0&.17  & 5&.14\,\, & $\pm$ 0&.17  & 6&.11\,\, &
 $\pm$ 0&.17 & 11&.9\,\, & $\pm$ 0&.7 & 20&\,\, & $\pm$ 2&  \\ 
16 & S  & 4&.84\,\, & $\pm$ 0&.12  & 4&.35\,\, & $\pm$ 0&.14 & 5&.32\,\, &
 $\pm$ 0&.14 & 15&.7\,\, & $\pm$ 0&.9 & 25&\,\, & $\pm$ 2&  \\  
29,30 & Brass & 5&.92\,\, & $\pm$  0&.11 & 9&.72\,\, & $\pm$ 0&.11 & 10&.69\,\,
& $\pm$ 0&.11 & 3&.9\,\, & $\pm$ 0&.1 & 14&.9\,\, & $\pm$ 0&.7  \\
 & (Cu+Zn) & & & & & & & & & & & & & & & & & & &\\
79 & Au & 6&.93\,\, & $\pm$  0&.08  & 13&.36\,\, & $\pm$ 0&.07 & 14&.33\,\, &
$\pm$ 0&.07 & 2&.2\,\, & $\pm$ 0&.1 & 10&.4\,\, &  $\pm$ 0& .3  \\
\end{tabular}
\label{tab2}
\end{centering}
\end{table}

\begin{table}[t]
\begin{centering}
 \caption{Values of S and t$_{0}$ for hyperfine targets, and their effect on
the systematic uncertainties of H and A.  The different results for H and A for
each target should be taken as a measure of the systematic uncertainties.
 }
\begin{tabular}{ccr@{}l@{}r@{}lr@{}l@{}r@{}lcr@{}l@{}r@{}lr@{}l@{}r@{}l} 
Z & Target & \multicolumn{4}{c}{S} & \multicolumn{4}{c}{t$_{0}$} & Type of &
\multicolumn{4}{c}{H} & \multicolumn{4}{c}{A} \\ 
 & & \multicolumn{4}{c}{(ns)} & \multicolumn{4}{c}{(ns)} & Error &
\multicolumn{4}{c}{($\mu$s$^{-1}$)} & \multicolumn{4}{c}{(unitless)} \\
 & & & & & & & & & & Functions & & & & & & & & \\ \hline 
9 & LiF  & 5&.38\,\, & $\pm$ 0&.18 (varied) & 5&.30\,\, & $\pm$ 0&.17 (varied)
 & 1 & 5&.6\,\, & $\pm$\,\,\, 0&.2 & 0&.288\,\, & $\pm$ 0&.006 \\ 
  &      & 4&.92\,\, & $\pm$ 0&.15 (varied) & 4&.81\,\, & $\pm$ 0&.15 (varied)
 &  2 & 5&.7\,\, & $\pm$\,\,\, 0&.3 & 0&.290\,\, & $\pm$ 0&.006  \\
  &      & 5&.31\,\, & $\pm$ 0&.20 (varied) & 5&.87\,\, & $\pm$ 0&.18 (varied)
 & 2 & 5&.6\,\, & $\pm$\,\,\, 0&.2 & 0&.287\,\, & $\pm$ 0&.006 \\
\hline
9 & (CF$_{2}$)$_{n}$ & 5&.03\,\, & $\pm$ 0&.31 (varied) & 5&.18\,\, & $\pm$
 0&.29 (varied) & 1 & 5&.2\,\, & $\pm$\,\,\, 0&.4 & 0&.33\,\, & $\pm$ 0&.01 \\
  &      & 4&.68\,\, & $\pm$ 0&.26 (varied) & 4&.59\,\, & $\pm$ 0&.27 (varied)
 & 2  & 5&.2\,\, & $\pm$\,\,\, 0&.4 & 0&.33\,\, & $\pm$ 0&.01 \\
  &      & 4&.89\,\, & $\pm$ 0&.36 (varied) & 5&.71\,\, &  $\pm$ 0&.31 (varied)
 & 2 & 5&.2\,\, & $\pm$\,\,\, 0&.4 & 0&.33\,\, & $\pm$ 0&.01 \\
\hline
11 & NaH  & 5&.0\,\, & $\pm$ 0&.2  (varied) & 5&.0\,\, &  $\pm$ 0&.1  (varied)
 & 1 & 26& &  $\pm$\,\,\, 4& & 0&.15\,\, & $\pm$ 0&.02 \\
   &      & 4&.31\,\, & $\pm$ 0&.22 (varied) & 4&.51\,\, & $\pm$ 0&.13 (varied)
 & 2 & 32& & $\pm$\,\,\, 4& & 0&.19\,\, & $\pm$ 0&.03 \\
   &      & 4&.94\,\, & $\pm$ 0&.16 (varied) & 5&.54\,\, & $\pm$ 0&.22 (varied)
 & 2 & 26& & $\pm$\,\,\, 5& & 0&.15\,\, & $\pm$ 0&.02 \\
\hline
11 & Na   & 4&.7\,\, &  $\pm$ 0&.1 (varied) & 4&.5\,\, & $\pm$ 0&.2  (varied) &
 1 & 39& & $\pm$\,\,\, 7& & 0&.18\,\, & $\pm$ 0&.04  \\
   &      & 4&.23\,\, &  $\pm$ 0&.13 (varied) & 3&.77\,\, & $\pm$ 0&.32
 (varied) & 2 & 49& & $\pm$  10& & 0&.24\,\, & $\pm$ 0&.06  \\
   &      & 5&.21\,\, & $\pm$ 0&.23 (varied) & 4&.75\,\, & $\pm$ 0&.14 (varied)
 & 2 & 34& & $\pm$\,\,\, 6& & 0&.15\,\, & $\pm$ 0&.03  \\
\hline
13 & Al  & 5&.13\,\, & $\pm$ 0&.13  (varied) & 3&.52\,\, & $\pm$ 0&.38 (varied)
 & 1 & 68& & $\pm$\,\,\, 9& & 0&.30\,\, & $\pm$ 0&.06 \\
   &     & 4&.63\,\, & $\pm$ 0&.13 (varied) & 2&.73\,\, & $\pm$ 0&.30 (varied)
 & 2 & 75& & $\pm$\,\,\, 7& & 0&.37\,\, & $\pm$ 0&.04 \\
   &     & 5&.21\,\, & $\pm$ 0&.23 (varied) & 4&.75\,\, & $\pm$ 0&.14 (varied)
 & 2 & 71& & $\pm$\,\,\, 9& & 0&.31\,\, & $\pm$ 0&.06 \\   
\hline
13 & LiAlH$_{4}$  & 5&.04\,\, & $\pm$ 0&.31  (varied) & 2&.37\,\, & $\pm$ 0&.45
 (varied) & 1 & 90& & $\pm$\,\,\, 6&  & 0&.50\,\, & $\pm$ 0&.04  \\
   &     & 4&.96\,\, & $\pm$ 0&.23 (varied) & 4&.66\,\, & $\pm$ 0&.85 (varied)
 & 2 & 53& & $\pm$ 25&  & 0&.22\,\, & $\pm$ 0&.13  \\
\hline 
15 & P  & 6&.51\,\, &  $\pm$ 0&.07  (varied) & 5&.7\,\, & $\pm$ 0&.23  (varied)
 & 1 & 44& & $\pm$\,\,\, 8& & 0&.12\,\, & $\pm$ 0&.03 \\
   &    & 5&.90\,\, &  $\pm$ 0&.07 (varied) & 4&.57\,\, & $\pm$ 0&.19 (varied)
 & 2 & 62& & $\pm$\,\,\, 6& & 0&.23\,\, & $\pm$ 0&.03  \\
   &    & 6&.73\,\, & $\pm$ 0&.11 (varied) & 5&.45\,\, & $\pm$ 0&.20 (varied) &
2 & 76& & $\pm$\,\,\, 5& & 0&.27\,\, & $\pm$ 0&.02  \\
\end{tabular}
\label{tab3}
\end{centering}
\end{table}

\begin{table}[t]
\begin{centering}
\caption{Muonic hyperfine transition rates in $\mu$s$^{-1}$.}
\begin{tabular}{cr@{}lr@{}l@{}r@{}lr@{}l@{}r@{}l}
Compound & \multicolumn{2}{c}{Theory\cite{r2}} &
\multicolumn{4}{c}{Previous Work} & \multicolumn{4}{c}{This experiment} \\
 \hline 
LiF & 5&.8  & 5&.8\, & $\pm$ 0&.8 \cite{r2} & 5&.6\, & $\pm$ 0&.2 \\
    & &     & 6&.3\, & $\pm$ 1&.8 \cite{r2} & & & & \\
Na\underline{F} & & & 4&.9\, & $\pm$ 1&.2 \cite{r4} & & & & \\
(CF$_{2}$)$_{n}$ & & & & & & & 5&.2\, & $\pm$ 0&.4 \\ \hline
Na & 14&  & 15&.5\, &  $\pm$ 1&.1 \cite{r5} & 38& & $\pm$ 9& \\
NaH & & & & & & & 28& & $\pm$ 5& \\
\underline{Na}F & && 8&.4\, &  $\pm$ 1&.9 \cite{r4} & & & & \\ \hline
Al & 41&  & 41& & $\pm$ 9 & \cite{r21} & 72& & $\pm$ 9& \\
LiAlH$_{4}$ & & & & & & & 89& & $\pm$ 10& \\ \hline
Red P & 58&  & \multicolumn{4}{c}{$\lambda_{h} \gg \lambda_{-}$
\tablenote{$\lambda_{-} \approx$ 1.1 $\mu$s$^{-1}$ }
 \cite{r4}} & 60& & $\pm$ 15& \\ 
 & & & \multicolumn{4}{c}{$\lambda_{-} \gg  \lambda_{h}$  \cite{r23}} & & & &
 \\ \hline
Li\underline{Cl} & 8&  & 6&.5\, & $\pm$ 0&.9 \cite{r4} & 14& & $\pm$ 27& \\
C\underline{Cl}$_{4}$ & & & & & & & 15&  &$\pm$ 29& \\ \hline
K & 22 & & & & & &   25& & $\pm$ 15& \\
\end{tabular}
\label{tab4}
\end{centering}
\end{table}

\begin{table}[t]
\begin{centering}
\caption{The capture rate asymmetry.  The theory values are for all captures.
This experiment is for neutron detection only (weighted by multiplicity).}
\begin{tabular}{cdddr@{}l@{}r@{}lr@{}l@{}r@{}l}  
Compound & BLYP\cite{r2} & Primakoff\cite{r2} & \"{U}berall\cite{r22} &
\multicolumn{4}{c}{Previous work \cite{r2}} &
\multicolumn{4}{c}{This Experiment} \\  \hline 
LiF & 0.24  & 0.36 & 0.36 & 0.&36\, & $\pm$ 0&.04 (n) & 0.&29\, & $\pm$ 0.&01\\
 & & & & 0.&25\, & $\pm$ 0&.04 ($\gamma$) & & & & \\
 & & & & 0.&29\, & $\pm$ 0&.02 (n$\gamma$) & & & & \\
(CF$_{2}$)$_{n}$ & 0.24  & 0.36  & 0.36 & & & &  & 0&.33\, & $\pm$ 0&.01 \\
 \hline
Na & 0.08  & & & & & & & 0&.17\, & $\pm$ 0&.05 \\
NaH & 0.08  & & & & & & & 0&.16\, & $\pm$ 0&.03 \\ \hline
Al & 0.09  & 0.14 & 0.22 & & & & & 0&.34\, & $\pm$ 0&.05 \\
LiAlH$_{4}$ & 0.09  & 0.14 & 0.22 & & & & & 0&.35\, & $\pm$ 0&.15 \\ \hline
Red P & 0.16  & 0.22 & 0.25 & & & & & 0&.2\, & $\pm$ 0&.1 \\ \hline
LiCl & -0.06 ($^{35}$Cl)  & -0.10 ($^{37}$Cl) & & & & & & -0&.03\, & $\pm$
 0&.06 \\
 & -0.06 ($^{37}$Cl) & & & & & & & & & & \\
CCl$_{4}$ & & & & & & & & -0&.03\, &  $\pm$ 0&.06 \\ \hline
K & -0.05 & -0.07 & & & & & & -0&.080\, & $\pm$ 0&.055\\

\end{tabular}
\label{tab5}
\end{centering}
\end{table}

\end{document}